\documentclass[amsfonts,amssymb,twocolumn,amsmath,showpacs,citesort]{revtex4-1}

\usepackage{mathrsfs}
\usepackage{amsfonts}
\usepackage{amstext}
\usepackage{amsmath}
\usepackage{amssymb}
\usepackage[dvips]{graphicx}
\def\qed{\leavevmode\unskip\penalty9999 \hbox{}\nobreak\hfill
     \quad\hbox{\leavevmode  \hbox to.77778em{%
               \hfil\vrule   \vbox to.675em%
               {\hrule width.6em\vfil\hrule}\vrule\hfil}}
     \par\vskip3pt}

\def\ra{\rangle}
\def\la{\langle}

\begin{document}
\title{Mixed maximally entangled states}
%

\author{Zong-Guo Li$^{1,2}$}
\author{Ming-Jing Zhao$^{3,4}$}
\author{Shao-Ming Fei$^{4}$}
\author{Heng Fan$^{2}$}
\author{W. M. Liu$^{2}$}

\affiliation{$^1$College of Science, Tianjin University of Technology, Tianjin 300384, China\\$^{2}$Beijing National Laboratory for Condensed Matter
Physics, Institute of Physics, Chinese Academy of Sciences, Beijing
100080, China\\
$^{3}$Max-Planck-Institute for Mathematics in the Sciences, 04103
Leipzig, Germany\\
$^{4}$Department of Mathematics, Capital Normal University,
Beijing 100037, China}

\begin{abstract}
We find that the mixed maximally entangled states exist and prove that the form of the mixed maximally entangled states is unique in terms of the entanglement of
formation. Moreover, even if the entanglement is quantified by other
entanglement measures, this conclusion is still proven right.  This result is a
supplementary to the generally accepted fact that all maximally
entangled states are pure. These states possess important properties
of the pure maximally entangled states, for example, these states
can be used as a resource for faithful teleportation and they can be
distinguished perfectly by local operations and classical
communication.

\end{abstract}
\pacs{03.67.Mn, 03.67.Hk, 03.65.Ud, 03.65.Yz, 89.70.+c} \maketitle

\section{Introduction}

Quantum entanglement is central both to the foundations of quantum
mechanics and to quantum information and computation
\cite{Vedral,Plenio,Horodecki,Benhelm}.
Its importance has been demonstrated in various applications such as
teleportation \cite{C.H.Be1,zhang,Modlawska,Ishizaka,Noh},
superdense coding \cite{C.H.Be2,Julio}, quantum computation
\cite{be3} and E91 protocol of quantum cryptography \cite{ekert},
etc. The {\it maximally} entangled states are especially important
for such quantum information processing tasks.
Experimentalists try continuously to improve the quality of the
entangled states by entanglement distillation and purification with
which one can create a small set of highly entangled states from a
large set of less entangled states (pure or mixed)
\cite{experiments}. Generally it is believed that the amount of
entanglement in a mixed state can be increased by purifying it
toward a pure maximally entangled state. That means the entanglement
is increased by increasing the purity of the mixed state. The reason
is simply because all known maximally entangled states are pure.
Actually it has been proved \cite{Cavaicanti} that all maximally
entangled states are pure in bipartite $d\otimes d$ systems. The
question is: Does there exist maximally entangled states which are
mixed states. And further, if a mixed maximally entangled state
exists, does it have any advantages and how to prepare it in a
physical system?


In this paper, we introduce a class of mixed states in $d\otimes
d'~(d'\!\geq\!2d)$ Hilbert space which are maximally entangled. We
prove that the form of the mixed maximally entangled states is
unique. On the one hand, this result is conceptually new since
really mixed maximally entangled states exist. On the other hand,
those mixed maximally entangled states which are proved to have a
unique form is actually equivalent to the pure maximally entangled
state tensor product an ancillary state.

For $d\otimes d'~(d'\!\geq\!2d)$ Hilbert space, a pure maximally
entangled state is
$|\phi\rangle\!\!=\!\!\sum_{i=1}^d\frac{1}{\sqrt{d}}|ii\rangle$.
If a mixed state has the same entanglement quantified by a certain
entanglement measure as this pure maximally entangled state, we call
it the mixed maximally entangled state. The entanglement measure can
be the well accepted entanglement of formation, or other
entanglement measures such as concurrence, distillable entanglement,
and the relative entropy of entanglement. The conclusion of this
paper does not change.

One key application of the pure maximally entangled state is perfect
teleportation. If a mixed state is used a quantum state will not be
teleported faithfully. However, this is only true for the case of
mixed state which is not maximally entangled. For mixed maximally
entangled states presented in this paper, we will show that these
states can be used as a resource for perfect teleportation. Also
those mixed maximally entangled states can be distinguished
perfectly by local operations and classical communication
\cite{WSHV,F,Hayashi,bose,Plenio1}. We also propose a preparation
scheme to create experimentally this kind of states.

\section{Mixed maximally entangled state}

We consider $d\otimes d'$ bipartite systems, assume $d'\geq d$
without loss of generality and take the well-accepted entanglement
measure, entanglement of formation \cite{wootters}. For a pure state
$|\psi\ra_{AB}$ shared by distant parties A and B, the entanglement
of formation $E$ is defined as the von Neumann entropy of either of
the two subsystems:
$E(|\psi\ra_{AB})=-\textrm{Tr}(\rho_{A(B)}\log_2\rho_{A(B)}),$ where
$\rho_{A(B)}=tr_{B(A)}(|\psi\ra_{AB}\la\psi|)$ is the reduced
density matrix; as for a mixed state $\rho_{AB}$, the entanglement
of formation is defined as the average entanglement of the pure
states $E(\rho)=\textrm{inf}\sum_ip_iE(|\psi_i\ra)$, minimized over
all pure state decompositions $\rho=\sum_ip_i|\psi_i\ra\la\psi_i|$,
$\sum_ip_i=1$. Hereafter we will use entanglement of formation to
quantify the entanglement in a state unless otherwise stated.

To show the existence of the mixed maximally entangled state, we
prove the following two lemmas:

\textbf{Lemma 1.} A pure $d\times d'$ bipartite state $|\psi\ra$ is
maximally entangled if and only if $E(|\psi\ra)=\log_2d$.

\emph{Proof.}
If a pure $d\times d'$ bipartite state $|\psi\ra$ is
maximally entangled, there must exist a local unitary operation $U_1\otimes U_2$ such that
\begin{equation}
(U_1\otimes U_2)|\psi\ra=\sum_{i=1}^d\frac{1}{\sqrt{d}}|ii\ra.
\end{equation}
Due to the invariance of entanglement of formation under local unitary
operation, we can get $E(|\psi\ra)=E((U_1\otimes U_2)|\psi\ra)=\log_2d$
in terms of the definition of entanglement of formation.

We next prove the necessary condition. Suppose $E(|\psi\ra)=\log_2d$. In terms of the Schmidt decomposition theorem, we can rewrite
this pure $d\times d'$ bipartite state $|\psi\ra$ as
\begin{equation}
\label{four lem11}
|\psi\ra=\sum_{i=1}^n \sqrt{x_i}|i_A\ra|i_B\ra,
\end{equation}
where $x_i$ is nonnegative and satisfies the condition $\sum_{i=1}^n x_i=1$,
$\{|i_A\ra\}$ and $\{|i_B\ra\}$ are orthonormal in Hilbert space $H_A$ and $H_B$, respectively.
In general, the Schmidt number $n$ satisfies $n\leq d$. In terms of the definition of entanglement of formation, we have
\begin{equation}
\label{max-eq3}
E(|\psi\ra)=-\sum_{i=1}^n x_i \log_2 x_i.
\end{equation}
Under the constraint condition $\sum_{i=1}^n x_i=1$, we can obtain the stable point, $x_i=\frac{1}{n}$, $\forall i$, by the method of Lagrange multipliers. By further analysis, the stable point is actually the maximum point of function (\ref{max-eq3}).
Therefore, the maximum value of Eq. (\ref{max-eq3}) $\log_2n$ is obtained if and only if all $x_i$s are equal, i.e.,  $x_i=\frac{1}{n}$, $\forall i$. As $E(|\psi\ra)=\log_2d$, $n=d$,
$x_i=\frac{1}{d}$, $\forall i$. Thus, this state $|\psi\ra$ is maximally entangled.
\qed

\textbf{Lemma 2.} $\rho$ is a mixed maximally entangled state with
respect to the entanglement of formation E, if and only if all pure-state
decompositions $\{p_i,\hspace{1mm}|\psi_i\ra \}$ of $\rho$ satisfy
the conditions that $E(|\psi_i\ra)=\log_2d$ for all $i$.

\emph{Proof.}
Firstly, we prove the sufficient condition. Assume $\rho$ is a mixed maximally entangled state with
respect to the entanglement formation E. Then, according to the definition of
entanglement of formation for mixed maximally entangled
states, we get its entanglement of formation, $E(\rho)=\log_2d$. As the entanglement of
formation of $\rho$ is defined as
\begin{equation}
\label{lem2}
E(\rho)=\min\sum_ip_iE(|\psi_i\ra)=\log_2d,
\end{equation}
where the minimum value is taken over all pure-state decompositions of $\rho$,
$\rho_{AB}=\sum_ip_i|\psi_i\ra\la\psi_i|$. Moreover, for an arbitrary pure $d\times d'$ state $|\psi\ra$,
its entanglement of formation isn't greater than $\log_2d$. Hence, all pure-state
decompositions $\{p_i,\hspace{1mm}|\psi_i\ra \}$ of $\rho$ must satisfy
the conditions that $E(|\psi_i\ra)=\log_2d$ for all $i$, otherwise the equation (\ref{lem2}) is violated.

If all pure-state
decompositions $\{p_i,\hspace{1mm}|\psi_i\ra \}$ of $\rho$ satisfy
the conditions that $E(|\psi_i\ra)=\log_2d$ for all $i$, it is easy to obtain
$E(\rho)=\log_2d$ in the light of the definition of entanglement of formation.
Therefore, the $d\times d'$ bipartite state $\rho$ is maximally entangled.
\qed

We also need the following lemma.

\textbf{Lemma 3.}
 $|\psi\ra$ is a $d\otimes d'$ pure maximally entangled state
if and only if for arbitrary given orthonormal complete basis
$\{|i_A\ra\}$ of the subsystem $A$, there exits an orthonormal basis
$\{|i_B\ra\}$ of the subsystem B such that $|\psi\ra$ can be written
in the following form,
\begin{eqnarray}
\label{puremaximallyentangled}
|\psi\ra=\frac{1}{\sqrt{d}}\sum_{i=1}^d|i_A\ra\otimes|i_B\ra.
\end{eqnarray}

\emph{Proof.} If a pure state $|\psi\ra$ is of the form
(\ref{puremaximallyentangled}), the entanglement of formation is
$E(|\psi\ra)=\log_2d$. In terms of the Lemma 1, the state $|\psi\ra$
is maximally entangled.

We next prove the sufficient condition. If the state $|\psi\ra$ is
maximally entangled, $|\psi\ra$ can always be written as
$|\psi\ra\!\!=\!\!\frac{1}{\sqrt{d}}\sum^d_{j=1}|V_j\ra\otimes|W_j\ra$
according to Schmidt decomposition theorem and Lemma 1, where
$|V_j\ra$ and $|W_j\ra$ are orthonormal basis vectors with respect
to subsystems $A$ and $B$, respectively. Due to the orthonormal
completeness of $\{|i_A\ra\}$ and $\{|V_j\ra\}$, the orthonormal
basis vector $|V_j\ra$ can also be represented as $
|V_j\ra\!\!=\!\!\sum_{i=1}^dU_{ji}|i_A\ra, $ where $U$
is a unitary matrix. The bipartite pure state $|\psi\ra$ then can be
written as $
|\psi\ra
\!\!=\!\!\frac{1}{\sqrt{d}}\sum^d_{i=1}|i_A\ra\otimes(\sum_{j=1}^dU_{ji}|W_j\ra)
\!\!=\!\!\frac{1}{\sqrt{d}}\sum_{i=1}^d|i_A\ra\otimes|i_B\ra,$ where
we have set $|i_B\ra\!\!=\!\!\sum_{i=1}^dU_{ji}|W_j\ra$ which is
just an orthonormal basis vector. \qed

The main result of this paper is the following£º

\textbf{Theorem 1.} A $d\otimes d'$ $(d'\geq2d)$ bipartite mixed
state $\rho$ is maximally entangled if and only if
\begin{eqnarray}
\label{mix} \rho\!=\!\sum_{m}
p_m|\psi_m\ra\la\psi_m|,\hspace{2mm}\sum_{m} p_m=1,
\end{eqnarray}
where $|\psi_m\ra\!\!=\!\!\frac{1}{\sqrt{d}}\sum_{i=1}^d|i\ra\otimes|i_m\ra$,
$\{|i\ra\}$ and $\{|i_m\ra\}$ are orthonormal bases of the
subsystems $A$ and $B$ respectively, satisfying $\la
j_n|i_m\ra=\delta_{ij}\delta_{nm}$.

{\em Proof.} If a mixed state $\rho$ has the pure-state
decomposition (\ref{mix}), which is the spectral decomposition of $\rho$, according to the method \cite{comm1},
a general decomposition $\{q_n,|w_n\ra\}$, $\rho=\sum_n q_n|w_n\ra\la w_n|$,
is given by $\sqrt{q_n}|w_n\ra\!\!=\!\!\sum_{m}U_{nm}\sqrt{p_m}|\psi_m\ra$,
$n\!\!=\!\!1,\ldots,l$. Here $U$ is an $l\times l$ unitary matrix, $l$ is greater than
or equal to the rank of $\rho$, and the following condition is satisfied, $\sum_m|U_{nm}|^2p_m/q_n=1$.
One can check that
\begin{eqnarray*}
\rho_A^{(n)}
&=&\textrm{Tr}_B|w_n\ra\la w_n|\\
&=&\textrm{Tr}_B(\sum_{m,m'} U_{nm}U_{nm'}^*\sqrt{p_m}\sqrt{p_{m'}}|\psi_m\ra\la\psi_{m'}|/q_n)\\
&=&\sum_{m,m'}\textrm{Tr}_B(|\psi_m\ra\la\psi_{m'}|)U_{nm}U_{nm'}^*\sqrt{p_m}\sqrt{p_{m'}}/q_n\\
&=&\frac{1}{d}\sum_{i=1}^d|i\ra\la i|\sum_{m,m'} U_{nm}U_{nm'}^*\sqrt{p_m}\sqrt{p_{m'}}\delta_{mm'}/q_n\\
&=&\frac{1}{d}I.
\end{eqnarray*}
Then, we obtain $E(|w_n\ra)\!\!=\!\!\log_2d$ for
an arbitrary pure-state decomposition $\{q_n,|w_n\ra\}$ of $\rho$.
Therefore from Lemma 1 and Lemma 2 we deduce that the mixed state
$\rho$ is maximally entangled.

If the $d\otimes d'$ bipartite mixed state $\rho$ is maximally
entangled, then in terms of Lemma 2 all states in the pure state
decomposition of $\rho$ are maximally entangled. Hence, the
eigenvectors of $\rho$ are also maximally entangled. In the light of
Lemma 3 the eigenvectors of $\rho$ are of the form,
$|v_m\ra\!\!=\!\!\sqrt{p_m}\frac{1}{\sqrt{d}}\sum_{i=1}^d|i\ra\otimes|\phi_{im}\ra$,
where $p_m$ is the $m$th eigenvalue, $\{|i\ra\}$ and
$\{|\phi_{im}\ra\}$ are orthonormal bases of subsystems $A$ and $B$
respectively, i.e., $\la
\phi_{jm}|\phi_{im}\ra\!\!=\!\!\delta_{ij}$. According to the method \cite{comm1}, then a general
decomposition $\{q_n,|u_n\ra\}$ of $\rho$ is given by,
\begin{eqnarray}
|u_n\ra&=&\sum^k_{m=1}U_{nm}|v_m\ra/\sqrt{q_n}\nonumber\\
&=&\frac{1}{\sqrt{d}}\sum_{i=1}^d|i\ra\otimes(\sum_{m=1}^k
U_{nm}\sqrt{p_m}|\phi_{im}\ra/\sqrt{q_n})\nonumber\\
&=&\frac{1}{\sqrt{d}}\sum_{i=1}^d|i\ra\otimes|\Phi_{in}\ra\quad n\!\!=\!\!1,\ldots,l
\label{max-eq1}
\end{eqnarray}
where $|\Phi_{in}\ra\!\!=\!\!\sum_{m=1}^k
U_{nm}\sqrt{p_m}|\phi_{im}\ra/\sqrt{q_n}$, $k$ is the rank of
$\rho_{AB}$, and $U$ is an $l\times l$ unitary matrix with
$l\!\!\geq\!\! k$. Because of the maximal entanglement in $|u_n\ra$,
the state $|\Phi_{in}\ra$ is orthonormal with respect to $i$  according to Lemma 3. Then we can get
\begin{eqnarray}
\label{max-eq2}
\la \Phi_{in}|\Phi_{jn}\ra&=&\sum_{m,m'}
U_{nm}U_{nm'}^*\sqrt{p_mp_{m'}}\la
\phi_{im'}|\phi_{jm}\ra/q_n\nonumber\\
&=&\delta_{ij}.
\end{eqnarray}

As long as we find an arbitrary $l\times l$ unitary matrix $U$ with
$l\!\geq\! k$, we will obtain a pure state decomposition $\{q_n,|u_n\ra\}$ of $\rho$ expressed as Eq.(\ref{max-eq1}). Furthermore, in terms of Lemma 2 and Lemma 3, the corresponding state $|\Phi_{in}\ra$ in Eq.(\ref{max-eq1}) must satisfy Eq.(\ref{max-eq2}) for any arbitrary $l\times l$ unitary matrix $U$. Due to the
arbitrariness of the unitary $U$, one can obtain that
$\la\phi_{im}|\phi_{im'}\ra\!\!=\!\!0$ for $m\neq m'$, and $\la
\phi_{im'}|\phi_{jm}\ra=0$ for $j\neq i$ and $m\neq m'$ by choosing
proper coefficients $U_{nm}$ \cite{coefficient}. This conclusion
gives rise to $\la
\phi_{jm'}|\phi_{im}\ra\!\!=\!\!\delta_{ij}\delta_{mm'}$, which
implies that the dimension $d'$ of subsystem $B$ must be greater or
equal to $kd$.

Therefore, a bipartite mixed state $\rho$ is maximally entangled if
and only if it has the form Eq. (\ref{mix}). \qed

From the theorem we see that, if the rank of a mixed
maximally entangled state $\rho$ of $d\otimes d'$ system is $k$,
$d'$ must be greater or equal to $kd$. For the case $k=1$, $\rho$
becomes a maximally entangled pure state. It is also evident that
there do not exist mixed maximally entangled states in $d\otimes d$
systems \cite{Cavaicanti}.

We now give an example of mixed maximally entangled state of
$2\otimes4$ bipartite systems, $
\rho\!\!=\!\!\frac{1}{2}(|\psi_1\ra\la\psi_1|+|\psi_2\ra\la\psi_2|),
$ where $|\psi_1\ra=\frac{1}{\sqrt{2}}(|00\ra+|11\ra)$ and
$|\psi_2\ra=\frac{1}{\sqrt{2}}(|02\ra+|13\ra)$ are both $2\times4$
maximally entangled pure states. Suppose that $\{q_i,|\phi_i\ra\}$
is an arbitrary pure-state decomposition of $\rho$,
$\rho=\sum_iq_i|\phi_i\ra\la\phi_i|$. Then there must exist a
unitary $U$ such that the general decomposition $|\phi_i\ra$ can be
given by $|\phi_i\ra=U_{i1}|\psi_1\ra+U_{i2}|\psi_2\ra$ with
$|U_{i1}|^2+|U_{i2}|^2=1$ \cite{comm1}. We have
$\rho_A^{(i)}=tr_B(|\phi_i\ra\la\phi_i|)=
\frac{1}{2}(|0\ra\la0|+|1\ra\la1|)$, and $E(|\phi_i\ra)=1$.
Therefore, $\rho_{AB}$ is a mixed maximally entangled state by the
Lemma 2.

{\em Remark.} As the classification and characterization of entanglement in multipartite states isn't fairly clear, this theorem is only valid for bipartite systems. If this mixed maximally entangled state is viewed as a multipartite state (e.g. the second Hilbert space is divided into two Hilbert spaces), then
it is equivalent to a pure maximally entangled state tensor product a mixed
state, $\rho\!=\!\sum_{m} p_m|\psi_m\ra\la\psi_m|=|\Phi ^+\rangle \langle \Phi ^+|\otimes \tilde {\rho }_a$, where $|\Phi ^+\rangle=\frac{1}{\sqrt{d}}\sum_{i=1}^d|i\ra\otimes|i\ra$, $\tilde {\rho }_a\equiv \sum p_m|m\rangle _a\langle m|$ with $|m\rangle$ orthonormal
eigenvectors corresponding to the eigenvalues $p_m$.
However, we would like to emphasize that our main result in this
paper is that for the first time, we identifies the unique form of
the mixed maximally entangled state. Physically the result can be
understood that particle $A$ which is a qudit entangles with particle
$B$ which is still a qudit but with $k$ different colors. Though we
may not know the exact color of particle $B$, but always $A$ and $B$
is maximally entangled.

In terms of entanglement of formation, we obtain the unique form of the mixed maximally entangled state. 
In addition, if entanglement measures $E(|\psi\rangle)$ can distinguish pure maximally entangled states from non-maximally entangled states, the Lemma 1 and 3 are easily proved to be valid by changing the value of entanglement measure for maximally
entangled states. On the basis of distinguishability for maximally entangled states, if the entanglement measures for mixed states are defined by the convex roof, the Lemma 2 must be true with the value of entanglement measure for maximally entangled states changed. In terms of Lemma 1, 2, 3, we can prove the Theorem 1 for such defined entanglement measures in a similar way. Therefore, as long as
entanglement measures defined by the
convex roof \cite{uhl} can distinguish maximally entangled states
from non-maximally entangled states, the Theorem 1 also holds for
these entanglement measures such as concurrence \cite{wootters}.

Moreover, for other
nonequivalent entanglement measures such as distillable entanglement
and the relative entropy of entanglement
\cite{vidal,bennett,vedral1,adam}, the theorem is also verified.

We will take the relative entropy of entanglement as an example. For
the state $
\rho=\frac{1}{4}[(|00\rangle+|11\rangle)(\langle00|+\langle11|)
+(|02\rangle+|13\rangle)(\langle02|+\langle13|)]$, we assume that there is a set
of local operations performed on the
second subsystem, a 4-dimensional system, and the corresponding operators are
expressed as,
$B_1=|0\rangle\langle0|+|1\rangle\langle1|$ and
$B_2=|2\rangle\langle2|+|3\rangle\langle3|$. Due to the monotonicity of
the relative entropy of entanglement under local operations and
classical communication (LOCC), that is, the relative entropy of entanglement
cannot increase under LOCC operations,
the relative entropy of this state
satisfies the following inequality,
$E_R(\rho)\geq \textrm{Tr}[(I\otimes B_1^\dag)\rho(I\otimes B_1)]
E_R(\frac{(I\otimes B_1)\rho(I\otimes B_1^\dag)}{\textrm{Tr}(I\otimes B_1^\dag)\rho(I\otimes B_1)})+
\textrm{Tr}[(I\otimes B_2)\rho(I\otimes B_2^\dag)]
E_R(\frac{(I\otimes B_2)\rho(I\otimes B_2^\dag)}{\textrm{Tr}(I\otimes B_2)\rho(I\otimes B_2^\dag)})
=1/2[E_R((|00\rangle+|11\rangle)/\sqrt2)+E_R((|02\rangle+|13\rangle)/\sqrt2)]$.
In the light of the theorem in the reference \cite{vedral2} and the
convexity of the relative entropy, we have $E_R(\rho)=1$ which means
this mixed state is maximally entangled.  In a similar way, for the
case with high dimension, this mixed state expressed in Eq. (\ref{mix}) of
our paper can be proved maximally entangled. Next we prove the mixed
maximally entangled state must be of the form of Eq. (\ref{mix}) for the
relative entropy. The quantity of entanglement in maximally
entangled states quantified by the relative entropy in $d\otimes
d'~(d'\geq d)$ Hilbert spaces
 is $\log_2d$. We suppose that
there exists a $d\otimes d'$ $(d'\geq d)$ mixed maximally entangled
state $\rho'$. Since the relative entropy is equal to the von
Neumann reduced entropy for pure states and is convex, we have the
following inequality:
$
E_R(\rho')\leq\sum_i p_iE_R(|\phi_i\rangle)=\sum_i
p_iE_N(|\phi_i\rangle),
$
where $E_N$ denotes the von Neumann reduced entropy and
$\{p_i,\phi_i\}$ is an arbitrary pure state decomposition of
$\rho'=\sum_ip_i|\phi_i\rangle\langle\phi_i|$, $\sum_ip_i=1$. Due to
the maximally entangled states $\rho'$, $E_R(\rho')=\log_2d$, the
inequality holds for any arbitrary pure-state decompositions of
$\rho'$, $\sum_i p_iE_N(|\phi_i\rangle)\geq\log_2d$. Therefore, this
inequality implies that the state $\rho'$ is also maximally
entangled for the entanglement of formation. In terms of the Theorem
1, the state $\rho'$ must be of the form of Eq. (\ref{mix}).

For the distillable entanglement $E_D$ which satisfies
$E_D\!\!\leq\!\!E_F$ with $E_F$ the entanglement of formation, we
can also prove, in a similar way, the mixed maximally entangled
states must be of the form of Eq. (\ref{mix}).

In fact, these mixed maximally entangled states can be used as
a resource for perfect teleportation and can be distinguished
perfectly by LOCC. In the following two sections, we provide a detailed protocol for perfect teleportation and perfect local distinguishability under LOCC.

\section{Teleportation with mixed maximally entangled state}

Suppose Alice
and Bob initially share a pair of particles, $A_2$ and $B$, in a
mixed maximally entangled state of $2d\otimes d$ system, $
\chi_{A_2B}=\frac{1}{2d}\sum_{i,j=0}^{d-1}(|i,i\ra\la
j,j|)+|d+i,i\ra\la d+j,j|). $ Alice wants to send an unknown state
of particle $A_1$,
$|\psi\ra_{A_1}=\sum_{i=0}^{d-1}\alpha_i|i\ra_{A_1}$, to Bob by
performing a complete von Neumann measurement on the joint system of
particles $A_1$ and $A_2$ and informing Bob the result of
measurement by classical communication.

We first define some operators so as to obtain the generalized Bell
states \cite{Sergio}. Let $h$ and $g$ be $d\times d$ matrices such
that $h|j\ra=|( j+1)\textrm{mod}\hspace{1mm}d\ra$,
$g|j\ra=\omega^j|j\ra$, with
$\omega=\textrm{exp}\{-2\textrm{i}\pi/d\}$. $d^2$ linear independent
$d\times d$ matrices are defined as $U_{st}=h^tg^s$. One can check
that $\{U_{st}\}$ satisfy the condition of \emph{basis of the
unitary operators} in the sense of \cite{werner}, i.e., $
\textrm{Tr}(U_{st}U^\dagger_{s't'})=d\delta_{tt'}\delta_{ss'}$, and
$U_{st}U^\dagger_{st}=I_{d\times d}$. Therefore, we can construct
$2d^2$ generalized Bell states:
\begin{eqnarray}
\label{mix1}
\begin{split}
|\Phi_{st}^1\ra\!\!&=&\!\!\frac{U_{st}\otimes
I}{\sqrt{d}}\sum_{i=0}^{d-1}|i,i\ra,\\
|\Phi_{st}^2\ra\!\!&=&\!\!\frac{U_{st}\otimes
I}{\sqrt{d}}\sum_{i=0}^{d-1}|i,d+i\ra.
\end{split}
\end{eqnarray}
$\{|\Phi_{st}^1\ra,~|\Phi_{st}^2\ra\}$ form a complete orthogonal
normalized basis of $d\otimes 2d$ system.

The initial state of the three particles, $A_1$, $A_2$ and $B$, can
be rewritten as: $ |\psi\ra_{A_1}\la\psi| \otimes \chi_{A_2B}
=\frac{1}{2d}\sum_{s,t,s',
t'}\Big{[}|\Phi_{st}^1\ra_{A_1A_2}\la\Phi_{s't'}^1|\otimes
U^\dagger_{st}|\psi\ra_B\la\psi|U_{s't'}
+|\Phi_{st}^2\ra_{A_1A_2}\la\Phi^2_{s't'}|\otimes
U^\dagger_{st}|\psi\ra_B\la\psi|U_{s't'}\Big{]}. $

To teleport the unknown state $|\psi\ra_{A_1}$ to Bob, Alice can
take a complete von Neumann measurement using the states
$\{|\Phi_{st}^i\ra\}$ on the joint system consisting of particles
$A_1$ and $A_2$. Then she announces her measurement result, $s$ and $t$, to Bob via classical communication. Bob can then transform
the state of his particle into $|\psi\ra_{A_1}$ by applying the
corresponding operation $U_{st}$. This means the mixed maximally
entangled state $\chi_{A_2B}$ can also be used as a resource for
perfect teleportation.

This is different from the usual concept in which the general mixed
state $\rho$ in $d\otimes d$ system can not be used to teleport an
unknown state $|\psi\ra$ in $d$-dimension with unit fidelity
\cite{Sergio,M.H}.

\section{Local distinguishability of the mixed maximally entangled
states}

Local distinguishability describes the property of states shared by
distant parties which are discriminated by only local operations and
classical communication. We start from a simple example: We have a
set of mixed maximally entangled states $\{ \rho(\Phi^{\pm }), \rho
(\Psi^{\pm })\} $, where $\rho(\Phi^{\pm })=\frac {1}{2}(|\Phi ^{\pm
}_1\rangle \langle \Phi ^{\pm }_1|+|\Phi ^{\pm }_2\rangle \langle
\Phi ^{\pm }_2|)$, $\rho(\Psi^{\pm })=\frac {1}{2}(|\Psi ^{\pm
}_1\rangle \langle \Psi ^{\pm }_1|+|\Psi ^{\pm }_2\rangle \langle
\Psi ^{\pm }_2|)$. Here states $|\Phi^{\pm }_1\rangle =\frac
{1}{\sqrt{2}}(|00\rangle \pm |11\rangle ), |\Psi^{\pm }_1\rangle
=\frac {1}{\sqrt{2}}(|01\rangle \pm |10\rangle )$ are four Bell
states, and correspondingly we denote $|\Phi^{\pm }_2\rangle =\frac
{1}{\sqrt{2}}(|02\rangle \pm |13\rangle ), |\Psi^{\pm }_1\rangle
=\frac {1}{\sqrt{2}}(|03\rangle \pm |12\rangle )$. We can find that
any two states from this set can be locally distinguished with
certainty, this is similar to the case of four Bell states.

We then present a general result about the local distinguishability
of the mixed maximally entangled states. We denote
\begin{equation}
\label{mix2}
\chi_{st}\!\!=\!\!\frac {1}{2}(|\Phi _{st}^1\rangle \langle |\Phi
_{st}^1|+|\Phi _{st}^2\rangle \langle |\Phi _{st}^2|),
\end{equation}
where $|\Phi_{st}^1\rangle$ and $|\Phi _{st}^2\rangle$ are defined in Eq. (\ref{mix1}),
then we have

\textbf{Theorem 2.} Any $l$ mixed maximally entangled states of the
set $\{\chi _{st}\} $ can be distinguished perfectly by local
operations and classical communication in case $l(l-1)\le 2d$, here
$d$ is prime.

The proof of this theorem is similar to the case of pure maximally
entangled states as in \cite{F}.

{\em Proof.}
According to Eq.(\ref{mix1}), the mixed maximally entangled states (\ref{mix2}) can
be rewritten as
\begin{eqnarray*}
\chi_{st}
&\!=\!&\frac {1}{2}(|\Phi_{st}^1\ra\la|\Phi_{st}^1|+|\Phi _{st}^2\ra\la\Phi_{st}^2|)\nonumber\\
&\!=\!&\frac{U_{st}\!\otimes\!I}{2d}\!\sum_{i,j=0}^{d-1}\!(|i,i\ra\la j,j|\!+\!|i,d\!+\!i\ra\la j,d\!+\!j|)(U^{\dag}_{st}\!\otimes\!I),
\end{eqnarray*}
where $U_{st}=h^sg^t$. $h$ and $g$ are defined in the last section, i.e.,
$h|j\ra=|(j+1)\textrm{mod}\hspace{1mm}d\ra$,
$g|j\ra=\omega^j|j\ra$, where $\omega=\exp\{2i\pi/d\}$.
Therefore, $h$ and $g$ can be expressed in the following
equation
\begin{equation*}
\label{four matrix1}
h\!\!=\!\!
\left(
   \begin{array}{ccccc}
   0 & 0 & \cdots & 0 & 1\\
   1 & 0 & \cdots & 0 & 0\\
   0 & 1 & \cdots & 0 & 0\\
   \vdots & \vdots & \vdots & \vdots & \vdots\\
   0 & 0 & \cdots & 1 & 0
   \end{array}
\right),\hspace{1mm}
g\!\!=\!\!
\left(
   \begin{array}{ccccc}
   1 & 0      & 0        & \cdots & 0\\
   0 & \omega & 0        & \cdots & 0\\
   0 & 0      & \omega^2 & \cdots & 0\\
   \vdots & \vdots & \vdots & \vdots & \vdots\\
   0 & 0      & 0        & \cdots & \omega^{d-1}
   \end{array}
\right),
\end{equation*}
and satisfy the conditions $gh=\omega h g$ and $g^{-1}h=\omega^{-1} h g^{-1}$.

For simplicity, we adopt the following denotation,
\begin{eqnarray}
\rho^+
=\frac{1}{2d}\sum_{i,j=0}^{d-1}\big(|i,i\ra\la j,j|+|i,d+i\ra\la j,d+j|\big).
\end{eqnarray}
Then, $\chi_{st}$ takes the following expression
$\chi_{st}=(U_{st}\otimes I)\rho^+(U^{\dag}_{st}\otimes I)$.
For an arbitrary $d\otimes d$ dimensional unitary matrix $V$
and  $2d\otimes 2d$ dimensional unitary matrix $V\oplus V$, we have
\begin{eqnarray}
\label{local3}
&&(I\!\otimes\! V)\sum_{i=0}^{d-1}|i,i\ra =(V^T\!\otimes\! I)\sum_{i=0}^{d-1}|i,i\ra,\\
\label{local4}
&&[I\!\otimes\!(V\!\oplus\! V)]\sum_{i=0}^{d-1}|i,d+i\ra\!\!=\!\!(V^T\!\otimes\! I)\sum_{i=0}^{d-1}|i,d+i\ra,
\end{eqnarray}
where $V^T$ denotes the transposition of $V$.

Suppose $l$ mixed maximally entangled states are denoted by $\{(h^{s_i}g^{t_i}\otimes I)\rho^+(h^{s_i}g^{t_i}\otimes I)^\dag\}_{i=1}^l$.
In terms of the equation $h|j\ra=|(j+1)\textrm{mod}\hspace{1mm}d\ra$, if the set $\{s_i\}_{i=1}^{l}$ has no equal $s_i$, we can locally distinguish these $l$ mixed maximally entangled states simply by performing
projecting-measurements in the computational basis $\{|i\ra\la i|\}$ on A side and $\{|i\ra\la i|+|d+i\ra\la d+i|\}$ on B
side respectively, and subsequently by a classical communication.

In general, to locally distinguish these states, we first
let A and B implement unitary operations $U$ and $V^T$, respectively.
This operation is equivalent to
the transformation $Uh^{s_i}g^{t_i}V\otimes I$ on $\rho^+$. We next
show that we can find these unitary operators $U$ and $V^T$ that can
transform these $l$ maximally entangled states to the set
$\{(h^{s'_i}g^{t'_i}\otimes I)\rho^+(h^{s'_i}g^{t'_i}\otimes I)^\dag\}_{i=1}^l$,
where there are no equal $s'_i$. As we have proved, this set can be distinguished locally.
Next we give these unitary operations.

We introduce $d$ generalized Hadamard transformations, $H_{\alpha}$, $(\alpha=0, 1, \cdots, d-1)$,
which are $d \otimes d$ dimensional unitary matrices with
\begin{eqnarray*}
(H_{\alpha})_{jk}\!=\!\omega^{-jk}\omega^{-\alpha m_k},
m_k\!=\!k\!+\!(k\!+\!1)\!+\!\cdots\!+\!(d\!-\!1).
\end{eqnarray*}
Deducing from the definition of $H_{\alpha}$ and the equation $g^{-1}h=\omega^{-1} h g^{-1}$, we have relations
$H_{\alpha}hH_{\alpha}^{\dag}=g^{-1}h^{\alpha}$,
$H_{\alpha}gH_{\alpha}^{\dag}=h,$ and
$H_{\alpha}h^{s_i}g^{t_i}H_{\alpha}^{\dag}=\Gamma h^{\alpha s_i+t_i}g^{-s_i}$, where the whole phase factor
$\Gamma=\omega^{-\alpha s_i(s_i+1)/2}$.
Firstly, let A and B implement unitary operations $H_{\alpha}$ and $H_{\alpha}^*\oplus H_{\alpha}^*$, respectively.
By applying Eqs. (\ref{local3}) and (\ref{local4}), the following equation holds
\begin{eqnarray}
&&[H_{\alpha}\otimes(H_{\alpha}^*\oplus H_{\alpha}^*)]\chi_{st}[H^\dag_{\alpha}\otimes (H_{\alpha}^*\oplus H_{\alpha}^*)^\dag]\nonumber\\
&=&(H_{\alpha}U_{st}H^\dag_{\alpha})\rho^+(H_{\alpha}U_{st}H^\dag_{\alpha})^\dag.
\end{eqnarray}
Given $l$ maximally entangled states corresponding to $\{h^{s_i}g^{t_i}\}_{i=1}^{l}$,
we can always transform them to the case $\{h^{s'_i}g^{t'_i}\}_{i=1}^{l}$, where the powers of $h$ are different, by
identity (do nothing) or $H_{\alpha}\otimes(H_{\alpha}^*\oplus H_{\alpha}^*)$, $(\alpha=0, 1, \cdots, d-1)$.
If not, then for each transformation at least two powers of $h$ are equal.
So at least we have $d+1$ equations altogether.
But different combinations between $l$ elements $\{h^{s_i}g^{t_i}\}_{i=1}^{l}$ is
$\tiny{\left(
   \begin{array}{c}
   l \\
   2
   \end{array}
\right)}=l(l-1)/2$, which is less than or equal to $d$.
This means two pairs, for example, $(s_0,t_0)$ and
$(s_1,t_1)$ without loss of generality, appear twice in two
different transformations, say $\alpha_0$ and $\alpha_1$,
that is $s'_0(\alpha_0)=s'_1(\alpha_0)$ and $s'_0(\alpha_1)=s'_1(\alpha_1)$.
Therefore, we obtain these equations,
\begin{eqnarray*}
\begin{split}
\alpha_0s_0+t_0\!&=&\!\!\alpha_0s_1+t_1\quad \quad (\textrm{mod}\hspace{1mm}d)\\
\alpha_1s_0+t_0&=&\alpha_1s_1+t_1\quad \quad (\textrm{mod}\hspace{1mm}d),
\end{split}
\end{eqnarray*}
Thus $(s_0,t_0)=(s_1,t_1)$, which contradicts our assumption
that these $l$ mixed maximally entangled states are orthogonal.
This completes our proof.
\qed

With mixed maximally entangled states, similar studies can be made
for other information processing tasks such as superdense coding,
quantum computation, cryptography, entanglement swapping and remote
state preparation.

\section{The experimental preparation}

The mixed maximally entangled state
may be realized in an NMR system. We may choose material $HC_2$ in
which the spin-1/2 $H$ and two isotope C13s, which are regarded as a
whole, provide a $2\otimes 4$ system \cite{du}. Suppose the system
$HC_2$ is in the initial state,
$|\psi\rangle=\sqrt{1/2}(|01\rangle-|10\rangle)$, and the two
isotope C13s interact with the environment---a spin-1/2 particle.
The spin-1/2 particle starts in the state $|0\rangle$ by applying an
external magnetic field with direction along z-axis. We also assume
that the interaction with the environment is described by the Hamiltonian,
$H\!\!=\!\!c(\sqrt{q_1}M_1\otimes
\sigma_z\!\!+\!\!\sqrt{q_2}M_2\otimes\sigma_x)$,
where $M_1=I_{4\times4}/\sqrt2$,
$M_2=(|0\ra\la2|+|1\ra\la3|+|3\ra\la1|+|2\ra\la0|)/\sqrt2$, and
$q_1+q_2=1$. $c$, $q_1$ and $q_2$ are real numbers and vary with the
intensity of magnetic field in the environment. Then we choose an
appropriate time and then obtain the Kraus operators $\sqrt{q_1}I\otimes M_1$
and $\sqrt{q_2}I\otimes M_2$ by performing a partial trace over the
environment. The final state becomes $\rho'=\sum_{i=1}^2q_i(I\otimes
M_i)|\psi\rangle\langle\psi|(I\otimes M_i^{\dagger})$, which is a
mixed maximally entangled state. However, the corresponding Hamiltonian between the spin-3/2 particle and an environment is not easy to manipulate. We
hope the scheme to prepare the mixed maximally entangled states may
help to provide some hints for the experimenters.

Due to the interactions with the environment in preparation and
transmission, the entangled pure states usually become mixed ones
and no longer entangled. However,
the entanglement evolution \cite{thomas,mt} for the mixed maximally entangled state  under the influence of local preparation channel shows that the output state is always a maximally entangled
state and still useful for many quantum information processing tasks
like perfect teleportation. This result is easily obtained from the entanglement evolution equation $\$(\rho
)=\sum_{i=1}^2 q_i(I\otimes M_i)\rho (I\otimes M_i^{\dagger})$, where 
$\rho
=(1-p)|\psi _1\rangle \langle \psi _1| +p |\psi _2\rangle \langle
\psi _2|$.

%
%


\section{Conclusions}

We have found a novel quantum state --- the mixed maximally
entangled state and prove that the form of the mixed maximally
entangled states is unique. A protocol is presented to teleport
faithfully an unknown $d$-dimensional state by resource of a  mixed
maximally entangled state in $d\otimes 2d$. Furthermore, it is shown
that any $l$ mixed maximally entangled states of the set
$\{(U_{st}\otimes I)\rho_{AB}(U^{\dagger}_{st}\otimes I)\}$ can be
discriminated perfectly by local operations and classical
communication. We also proposed a scheme to prepare these states in
an NMR physical system.

This work was supported by NSFC under Grants No. 10674162, 10874235,
10934010, 11047015 and 60978019, the NKBRSFC under Grants No. 2006CB921400,
2009CB930704, 2010CB922904, and 2011CB921502.


\begin{thebibliography}{18}
\bibitem{comm1} According to the spectral decomposition theorem, we can rewrite a state $\rho$
as $\rho=\sum_{i=1}^n p_i|v_i\ra\la v_i|$, where $|v_i\ra$ are orthonormal
eigenvectors corresponding to the nonzero eigenvalues $p_i$, $n$ is the rank of $\rho$. Then a general
pure-state decomposition $\{|w_i\ra\}$ of $\rho$ is given by $|w_i\ra=\sum_{j=1}^nU_{ij}\sqrt{p_j}|v_j\ra$, $i=1\cdots m$.
Here $U$ is an $m\times m$ unitary matrix, and $m$ is greater than
or equal to $n$. According to the unitarity of $U$, $\sum_i|w_i\rangle\langle w_i|=\rho$ can be verified.
The states $|w_i\ra$ are automatically subnormalized
so that $\la w_i|w_i\ra$ is equal to the probability of $|w_i\ra$ in the
decomposition.
\bibitem{coefficient}
One can always find a unitary matrix $U$ such that $U_{nn_1}=1/\sqrt{2},\hspace{2mm}U_{nn_2}=\textrm{i}/\sqrt{2}\hspace{2mm}(n_1\neq n_2)$ for a fixed $n$ (The elements with $m \neq n_1,n_2$ are zero). Then we have $\la\phi_{in_1}|\phi_{jn_2}\ra-\la\phi_{in_2}|\phi_{jn_1}\ra=0$. We can also find one other unitary matrix $U$ such that $U_{nn_1}=1/\sqrt{2},\hspace{2mm}U_{nn_2}=1/\sqrt{2} \hspace{2mm}(n_1\neq n_2)$. Correspondingly, the equation
$\la\phi_{in_1}|\phi_{jn_2}\ra+\la\phi_{in_2}|\phi_{jn_1}\ra=0$ is
satisfied. Hence $\la\phi_{in_2}|\phi_{jn_1}\ra=0$. Then it is easy to obtain
$\la\phi_{im}|\phi_{jm'}\ra=\delta_{ij}\delta_{mm'}$.
\end{thebibliography}
\end{document}